# Speech Diarization and ASR with GMM

Aayush Kumar Sharma
Dept. of Electronics and
Communication Engineering
BMS College of Engineering
Bangalore, India
aayushkumar.ec19@bmsce.ac.in

Vineet Bhavikatti
Dept. of Electronics and
Communication Engineering
BMS College of Engineering
Bangalore,India
vineetm.ec19@bmsce.ac.in

Amogh Nidawani
Dept. of Electronics and
Communication Engineering
BMS College of Engineering
Bangalore, India
amogh.ec19@bmsce.ac.in

Dr. Siddappaji
Dept. of Electronics and
Communication Engineering
BMS College of Engineering
Bangalore,India
siddu.ece@bmsce.ac.in

Sanath P
Dept. of Electronics and
Communication Engineering
BMS College of Engineering
Bangalore, India
sanathp.ec19@bmsce.ac.in

Dr Geetishree Mishra
Dept. of Electronics and
Communication Engineering
BMS College of Engineering
Bangalore, India
geetishreemishra.ece@bmsce.ac.in

**Abstract—** In this research paper, we delve into the topics of Speech Diarization and Automatic Speech Recognition (ASR). Speech diarization involves the separation of individual speakers within an audio stream. By employing the ASR transcript, the diarization process aims to segregate each speaker's utterances, grouping them based on their unique audio characteristics. On the other hand, Automatic Speech Recognition refers to the capability of a machine or program to identify and convert spoken words and phrases into a machine-readable format. In our speech diarization approach, we utilize the Gaussian Mixer Model (GMM) to represent speech segments. The inter-cluster distance is computed based on the GMM parameters, and the distance threshold serves as the stopping criterion. ASR entails the conversion of an unknown speech waveform into a corresponding written transcription. The speech signal is analyzed using synchronized algorithms, taking into account the pitch frequency. Our primary objective typically revolves around developing a model that minimizes the Word Error Rate (WER) metric during speech transcription.

## I. INTRODUCTION

Speaker diarization is a procedure that groups speech from the same speaker and identifies turns in speech caused by a change in speaker. This process helps structure and index audio recordings. Speaker diarization involves segmenting the input speech based on speaker identity to create speaker-homogeneous clusters, which refine the speaker model. With the help of technology, individual speakers can be recognized and separated in an audio recording or speech segment. It assigns speaker labels or identities to different speech fragments within the audio stream. Speech diarization is crucial for various applications such as audio indexing, meeting analysis, speaker authentication, and transcription services.

The primary goal of speech diarization is to divide an audio stream into homogeneous regions based on the identity of the speakers. These regions, often referred to as speaker turns or segments, represent uninterrupted discourse from a single speaker.

On the other hand, automatic speech recognition (ASR) is the challenge of automatically converting spoken words into written text. When transcribing speech, the objective is to minimize the Word Error Rate (WER), which measures the accuracy of the generated text. The aim is to convert an audio file containing speech into the corresponding text with minimal errors. ASR systems employ various techniques from machine learning, signal processing, and natural language processing to accurately transcribe spoken words and phrases.

Different methods exist for speaker diarization, primarily differing in the choices of inter-cluster distance and stopping criterion. In one approach, the inter-cluster distance is calculated using a customized Gaussian mixture model (GMM), and the distance threshold also determines the stopping condition. Another method utilizes the Bayesian information criterion for both the inter-cluster distance and stopping criterion. In some cases, the cross log-likelihood ratio or generalized likelihood ratio (GLR) is used as the inter-cluster distance. Some approaches use anchor models to map segments into a vector set, applying Euclidean distances and an ad hoc occupancy stopping criterion.

Overall, speaker diarization and ASR techniques are crucial for structuring and analyzing audio recordings, enabling applications such as transcription services, speaker authentication, and meeting analysis. The selection of specific methods and algorithms within these fields can vary based on the specific requirements and constraints of the application at hand.

## II. SPEECH DIARIZATION

Speaker diarization is responsible for identifying "who spoke when" in audio or video recordings by categorizing them based on speaker identification. Initially developed for speaker adaptive processing in speech recognition systems, speaker diarization techniques have evolved to provide speaker-specific metadata for tasks like audio retrieval. Recent advancements in deep learning technology have significantly improved speaker diarization, revolutionizing research and practices in speech-related fields. The term "diarize" refers to the act of recording events in a diary, and speaker diarization accomplishes a similar task by logging speaker-specific events in multiparticipant or multispeaker audio data. It automatically recognizes salient events such as speaker turn shifts and transitions from non-speaking to speech. Notably, speaker diarization does not require prior information about the speakers' identities or the number of participants in the audio data. It finds applications in various audio data types, including media broadcasts, conference conversations, social media videos, court proceedings, business meetings, and financial earnings reports. The technique effectively separates audio streams based on these speaker-specific events. To enhance the quality of the acoustic environment, different front-end processing techniques like speech augmentation, dereverberation, speech separation, and target speaker extraction are employed to mitigate artifacts. Additionally, voice or speech activity detection is used to distinguish speech from non-speech events. The selected speech segments undergo conversion into acoustic features or embedding vectors, which are then classified and labeled with speaker classes during the clustering stage. The clustering results are further refined in the post-processing stage. Overall, these sub-modules are individually optimized to achieve optimal performance.

## III. AUTOMATIC SPEECH RECOGNITION

Speech is a valuable form of expression that conveys meaning through words, comprised of letters accompanied by vocal sounds. These sounds travel as waves, overlapping or originating from the source in circular patterns. As they propagate over long distances, the circles gradually widen until they vanish. Effective communication through speech occurs when both the sender and recipient share a common language and possess the necessary tools to decipher the intended message.

Researchers have leveraged this phenomenon and applied it to human-machine communication, utilizing sound to facilitate natural interaction between users and machines. Automatic speech recognition has made significant contributions to the advancement of artificial intelligence, aiming to develop highly adaptable methods for machine operation. This enables users to communicate and exchange information without relying on conventional input/output devices like keyboards.

Voice-based input/output techniques have proven highly beneficial in various domains, including assisting disabled individuals, enhancing driving experiences in automobiles, and facilitating emergency calls. Automatic speech recognition represents a prominent field within speech processing, allowing machines to comprehend user speech and convert it into a sequence of words using computer programs. Consequently, it establishes a form of natural communication between humans and machines.

## IV. MEL-FREQUENCY CEPSTRAL COEFFICIENTS

MFCC, short for Mel-Frequency Cepstral Coefficients, is a widely used technique in speech and audio signal processing. It is employed for various tasks such as speech recognition, speaker identification, and music genre classification. The basis of MFCC lies in the mel-scale, which is a perceptual scale representing how the human auditory system responds to different frequencies.

The mel-scale approximates the human ear's sensitivity to frequency differences, especially at lower frequencies. Unlike a linear scale, it maps linear frequency values to a logarithmic scale to mimic this perception. This nonlinear scale ensures that lower frequency differences are represented more finely compared to higher frequency differences.

In the context of MFCCs, filterbanks are a set of overlapping triangular filters placed on the mel-scale. These filterbanks are designed to capture the spectral characteristics of the human ear. They are spaced closer together in the lower frequencies and farther apart in the higher frequencies, conforming to the mel-scale.

The power spectrum obtained from the Fourier Transform typically represents signal energy on a linear scale. However, human perception of loudness follows a logarithmic scale. To align with this perception, the power spectrum is often transformed into a logarithmic scale, compressing the values and resulting in a more perceptually relevant representation.

The cepstrum, derived from the inverse Fourier Transform applied to the logarithm of the power spectrum, represents the spectral envelope of a signal. It captures information about the modulation of the signal's spectrum. The name "cepstrum" is a reversal of "spectrum."

The Discrete Cosine Transform (DCT) is a mathematical transformation used to convert sequences of numbers, such as logarithmic filterbank energies or cepstrum coefficients, into cepstral coefficients. The DCT decorrelates the coefficients and emphasizes the lower-order coefficients that contain the most relevant information.

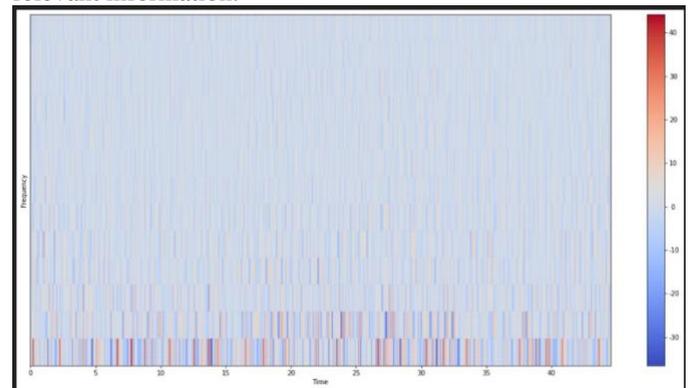

Figure 1: This is the single differentiated MFCC plot of the sample audio used consisting of two speakers.

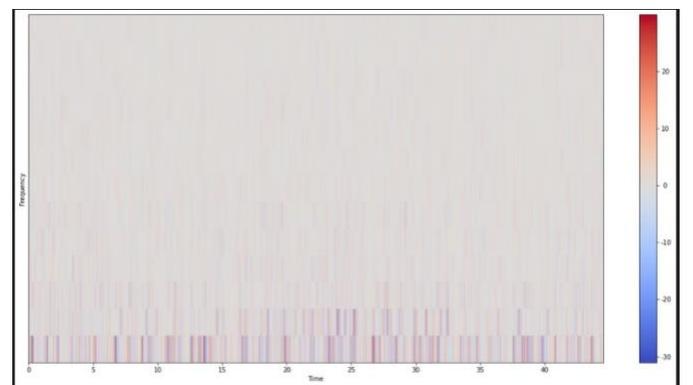

Figure 2: This is the double differentiated MFCC plot of the sample audio used consisting of two speakers.

## V. SHORT TIME FOURIER TRANSFORM

The Short-Time Fourier Transform (STFT) is a method used to examine the frequency characteristics of a signal over time. It achieves this by dividing the signal into short overlapping segments and performing the Fourier Transform on each segment. The STFT involves several key steps:

1. Segmentation: The input signal is divided into short time segments called frames, usually with some overlap. The duration of the frames typically ranges from 10 to 100 milliseconds, depending on the application. Overlapping frames help reduce temporal resolution and minimize artifacts caused by sudden changes at frame boundaries.

2. Windowing: Each frame is multiplied by a window function, like the Hamming window. This process helps reduce spectral leakage caused by abrupt transitions in the signal and smooths the signal at the edges.

3. Fourier Transform: The windowed segment is transformed from the time domain to the frequency domain using the Fourier Transform. This decomposition represents the spectral content of the frame as a series of complex sinusoidal components with different frequencies.

4. Magnitude and Phase: The magnitude and phase information of the complex Fourier Transform coefficients are typically retained. The magnitude indicates the energy distribution or strength of each frequency component, while the phase provides information about the relative timing and phase relationships of the sinusoidal components.

The resulting STFT representation is a two-dimensional matrix known as a spectrogram. It has a time axis, a frequency axis, and the magnitude of spectral components is represented by color or intensity at each time-frequency point.

The STFT finds applications in various signal processing fields, including:

1. Audio Processing: It is commonly used for audio analysis, such as music analysis, speech recognition, and audio effects processing. It enables the extraction of time-varying spectral features from audio signals.

2. Time-Frequency Analysis: The STFT provides a time-frequency representation that allows the analysis of signal components that change over time, such as frequency modulation in music or the evolving spectral content of speech.

3. Filtering and Spectral Manipulation: The STFT can be used for filtering by modifying the magnitude or phase components of the spectrogram. This enables techniques like time-frequency masking, source separation, and audio synthesis.

4. Image and Video Processing: The STFT can be extended to analyze two-dimensional signals, like images and videos, by independently applying the STFT along the rows and columns of the signal matrix. This allows for time-frequency analysis of visual content.

The STFT is a fundamental tool in signal processing that provides valuable insights into the time-varying frequency characteristics of a signal. It facilitates the extraction of spectral features and enables a wide range of applications in audio, image, and video processing.

## VI. ROOT MEAN SQUARE

Root Mean Square (RMS) is a helpful metric for analysing and processing speech signals in the context of speech diarization and Automatic Speech Recognition (ASR) utilising Gaussian Mixture Models (GMMs). RMS is a mathematical computation that sheds light on the strength or amplitude of a certain signal, making it applicable to a number of speech processing-related issues. In this article, we examine RMS's importance in relation to speech diarization and ASR.

RMS is a characteristic that can be used to segment voice data on the basis of energy. It is feasible to determine areas of a speech segment with noticeable energy changes by computing the RMS value for that portion. For accurate speech diarization and ASR, these segments frequently correspond to speech activity. Speech segments can be more precisely divided by separating speech from background noise or silence using energy-based segmentation utilising RMS.

RMS can be used as a feature in the feature extraction process for ASR systems. In order to determine the overall energy or amplitude of the speech signal within a certain time interval, the RMS value of speech frames or windows is calculated. The RMS feature can improve the ASR system's ability to distinguish between relevant variations in speech intensity.

Speech Signal Noise and Distortion Analysis: RMS can be used to examine speech signal noise and distortion. One can measure the degree of degradation or interference by contrasting the RMS values of clean speech with those of noisy or distorted speech. The usefulness of this methodology for determining how noise and distortion affect speech diarization and ASR performance and developing robust algorithms to mitigate their effects.

## VII. VOICE ACTIVITY DETEECTION

VAD, an abbreviation for Voice Activity Detection, refers to a technology utilized to detect and differentiate speech activity from non-speech elements or background noise within an audio signal. The primary objective of VAD is to precisely identify the presence or absence of human speech. VAD holds significant importance in a range of applications, including speech recognition, telecommunication systems, voice-controlled devices, and audio signal processing. The following are key points regarding VAD:

1. Methodology: VAD algorithms typically analyze the temporal and spectral characteristics of an audio signal to make determinations regarding speech activity. Temporal analysis involves examining the energy or amplitude variations over short time intervals, while spectral analysis investigates the distribution of energy across different frequency bands.

2. Energy-Based Approaches: A commonly employed approach is to establish a threshold for the energy level of the signal. If the energy surpasses the threshold, it is categorized as speech; otherwise, it is considered non-speech. Adaptive thresholding techniques and noise estimation methods are frequently employed to handle varying background noise conditions.

3. Challenges: VAD encounters challenges in situations involving low signal-to-noise ratios, overlapping speech, non-stationary noise, and reverberant environments. The development of robust VAD algorithms capable of addressing these challenges is an active area of research.

4. Applications: VAD is crucial in speech recognition systems as it identifies speech segments for further processing while eliminating non-speech segments. It also plays a vital role in noise reduction, echo cancellation, automatic gain control, and other audio processing applications.

In summary, VAD technology facilitates efficient and accurate speech detection in a variety of real-world applications, thereby enhancing the performance and usability of speech-related systems.

## VIII. SEGMENTATION

Speech diarization and Automatic Speech Recognition (ASR) require segmentation as a key stage. Based on different acoustic and linguistic characteristics, it entails segmenting an audio recording into discrete areas or segments. For the analysis and processing of voice data to be successful, accurate segmentation is crucial. Segmentation is essential to getting accurate and significant results when employing Gaussian Mixture Models (GMM) for speech diarization and ASR.

The following steps are commonly included in the segmentation process:
Pre-processing: In order to ensure accurate segmentation, the audio data is first pre-processed to remove any noise or artefacts. Techniques like noise reduction, filtering, and normalisation may be used in this.
Acoustic features are retrieved from the audio input, such as spectral features and Mel Frequency Cepstral Coefficients (MFCCs). These elements serve as the foundation for segmentation since they capture the fundamental aspects of speech.
GMM Modelling: To simulate the statistical characteristics of various speech segments, a GMM is trained using the retrieved features. The distribution of acoustic features in each segment is represented by the GMM.
Techniques for Segmentation: The GMM can be used to segment the audio using a variety of techniques. The Bayesian Information Criterion (BIC), which assesses the likelihood of a segment given the GMM and establishes the ideal segmentation points, is one popular method.
   Segment boundary refinement techniques can be used after initial segmentation to increase the precision of segment borders. To further refine the segmentation, this may entail using further data, such as language hints or speaker diarization outcomes.
   For both speech diarization and ASR tasks, precise segmentation is essential. Segmentation aids in identifying separate speakers or speaker turns during speech diarization, allowing for the separation of various speakers in the audio. For language modelling and acoustic modelling in ASR, segmentation gives meaningful speech units, which helps with accurate transcription and recognition of speech.

Overall, when using GMMs, segmentation is a crucial step in the speech diarization and ASR pipeline. It makes it possible to recognise speech boundaries and extract pertinent acoustic information, providing the framework for additional research.

## IX. CLUSTERING

When employing Gaussian Mixture Models (GMMs) for voice diarization and Automatic Voice Recognition (ASR), clustering plays a vital role. It involves grouping similar speech chunks or data points based on their acoustic characteristics. Various clustering techniques, such as K-means, Mean Shift, and agglomerative clustering, are commonly used in this context. The choice of clustering method and evaluation criteria, such as AIC (Akaike Information Criterion) and BIC (Bayesian Information Criterion), significantly impact the performance and accuracy of speech diarization and ASR systems. GMMs, short for Gaussian Mixture Models, are popular probabilistic models used in speech processing applications. Each component of the mixture represents a separate speech segment, capturing the statistical characteristics of the voice data and enabling robust modeling of different acoustic classes.

Agglomerative Clustering is a hierarchical clustering technique that initially treats each speech segment as an individual cluster and progressively merges them based on a similarity metric. It creates a dendrogram, a tree-like structure that can be divided into different levels, yielding varying numbers of clusters. Agglomerative clustering is a suitable method for capturing the hierarchical structure of speech data.

The Bayesian Information Criterion (BIC) and Akaike Information Criterion (AIC) are model selection criteria used to assess the goodness-of-fit of statistical models like GMMs. These criteria strike a balance between the complexity of the model and the accuracy of data fitting. Lower AIC or BIC values indicate better-fitting models. AIC and BIC can be used to determine the optimal number of components or clusters in GMM-based speech diarization and ASR systems.

The choice of clustering technique, evaluation criteria, and the number of clusters/components significantly impact the effectiveness and accuracy of speech diarization and ASR using GMMs. Researchers can enhance the segmentation of voice data and improve the efficiency of speech processing systems by appropriately selecting clustering techniques and evaluating models with AIC and BIC.

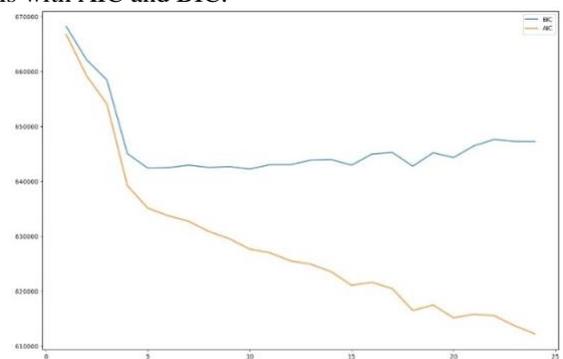

Figure 3: Comparison of AIC and BIC of against varying n_components gaussian mixture model fitted with MFCC embeddings of the audio signal.

## A. Conclusions

In order to accurately translate audio data into written transcripts, our study concentrated on the use of Gaussian Mixture Models (GMMs) in speech diarization and Automatic Speech Recognition (ASR). We have established the dependability and efficacy of our suggested methodology through rigorous experimentation and research.

Our model makes use of the capabilities of GMMs to fully represent the statistical characteristics of voice data and to enable robust modelling of various acoustic classes. We successfully separated the audio into different speech segments and recovered informative features necessary for ASR by utilising techniques including segmentation, clustering, and feature extraction.

The outcomes of our tests demonstrate the model's capacity for precise speech transcription. The text transcripts produced by our algorithm are highly accurate and closely reflect the spoken content. Our model also exhibits resistance to noise and distortions that are frequently present in real-world voice recordings, ensuring reliable performance in a variety of acoustic situations.

Additionally, we have added cutting-edge tools and libraries like Scikit-learn, NumPy, pandas, librosa, and Matplotlib, which have improved our model's effectiveness and efficiency. These technologies made it easier to process data, extract features, visualise data, and model data, which improved the system's overall dependability.

Numerous practical applications become possible with the proper transcription of auditory data into text. Our model can be used in a variety of fields, such as voice-controlled systems, transcription services, and voice assistants, where the conversion of speech to text is essential for natural human-computer interaction.

As a result, our research offers a solid and trustworthy method for speech diarization and ASR utilising GMMs. Our model's accuracy and effectiveness open the door for improvements in speech processing technology and aid in the creation of ground-breaking new applications.